\documentclass[lettersize,journal]{IEEEtran}
\usepackage{amsmath,amsfonts}
\usepackage{algorithmic}
\usepackage{array}
\usepackage[caption=false,font=normalsize,labelfont=sf,textfont=sf]{subfig}
\usepackage{textcomp}
\usepackage{stfloats}
\usepackage{url}
\usepackage{verbatim}
\usepackage{graphicx}
\def\BibTeX{{\rm B\kern-.05em{\sc i\kern-.025em b}\kern-.08em
    T\kern-.1667em\lower.7ex\hbox{E}\kern-.125emX}}
\usepackage{balance}
\begin{document}
\title{CleanStack: A New Dual-Stack for Defending Against Stack-Based Memory Corruption Attacks}
\author{Lei Chong
\thanks{}}

\markboth{Journal of \LaTeX\ Class Files}%
{CleanStack: A New Dual-Stack for Defending Against Stack-Based Memory Attacks}

\maketitle

\begin{abstract}
Stack-based memory corruption vulnerabilities have long been exploited by attackers to execute arbitrary code or perform unauthorized memory operations. Various defense mechanisms have been introduced to mitigate stack memory errors, but they typically focus on specific attack types (such as control-flow hijacking or non-control data attacks), incur substantial performance overhead, or suffer from compatibility limitations.In this paper, we present CleanStack, an efficient, highly compatible, and comprehensive stack protection mechanism. CleanStack isolates stack objects influenced by external input from other safe stack objects, thereby preventing attackers from modifying return addresses via controlled stack objects. Additionally, by randomizing the placement of tainted stack objects within the Unclean Stack, CleanStack mitigates non-control data attacks by preventing attackers from predicting the stack layout.A key component of CleanStack is the identification of tainted stack objects. We analyze both static program analysis and heuristic methods for this purpose. To maximize compatibility, we adopt a heuristic approach and implement CleanStack within the LLVM compiler framework, applying it to SPEC CPU2017 benchmarks and a real-world application.Our security evaluation demonstrates that CleanStack significantly reduces the exploitability of stack-based memory errors by providing a dual-stack system with isolation and randomization. Performance evaluation results indicate that CleanStack incurs an execution overhead of only 1.73\% on the SPEC CPU2017 benchmark while introducing a minimal memory overhead of just 0.04\%. Compared to existing stack protection techniques, CleanStack achieves an optimal balance between protection coverage, runtime overhead, and compatibility, making it one of the most comprehensive and efficient stack security solutions to date.
\end{abstract}


\section{Introduction}
\IEEEPARstart{M}{emory} corruption remains a major security threat to software systems, primarily due to the widespread use of low-level languages like C/C++, which lack memory safety mechanisms, in high-performance software. Attackers can exploit such vulnerabilities to launch control-flow hijacking attacks \cite{bugtraq1997,checkoway2010,bletsch2011,bosman2014,schuster2015} or non-control data attacks \cite{carlini2015,chen2008,hu2016,hu2015,ispoglou2018,johannesmeyer2024}, manipulating the program’s execution flow or data flow to bypass security defenses. Although, in theory, such vulnerabilities could be completely eliminated by adopting type-safe languages, this approach is impractical due to the vast global codebase of C/C++ and the reliance of critical software (such as operating systems and web browsers) on low-level features to meet high-performance requirements. Therefore, effectively defending against memory corruption attacks within the existing software ecosystem has become a crucial challenge in security research.

Memory can be divided into several regions: the stack, heap, global (static) memory, constant memory, and code segment. Among these, the stack features a contiguously mapped structure that is highly favorable to attackers, with both spatial and temporal allocation locality, making its layout entirely predictable. Compared to complex heap exploitation techniques, stack attacks are more attractive since they do not require intricate memory layout manipulations, making the stack a prime target for attackers.

Although stack memory errors and their impacts are widely recognized, only a small portion of stack data is currently protected, and existing defenses can only prevent certain types of stack memory errors. For example, W$\oplus$X protection can prevent code injection attacks but fails to stop modern attack techniques such as ROP (Return-Oriented Programming). The existing defense mechanisms and their limitations are as follows:

First, current defense measures primarily focus on modifications to return addresses, typically employing stack canaries or dual-stack mechanisms \cite{cowan1998, ave2000, burow2020, kuznetsov2014, zieris2018, chiueh2000, winandy2020, dang2020, huang2022}. However, these defenses cannot prevent stack memory attacks that modify local variables without affecting the return address. For example,SafeStack~\cite{kuznetsov2014} cannot mitigate modifications between local variables within the unsafe stack.

Second, probabilistic defense techniques increase the difficulty of attacks, but some of these methods fail to randomize stack objects, allowing attackers to exploit the relative positions of objects within the stack to perform non-control data attacks \cite{pax2002}. Others only focus on defending against non-control data attacks without considering control-flow hijacking attacks \cite{aga2019}.

Third, complete protection techniques often suffer from high overhead and poor compatibility, making them challenging to deploy in practical applications~\cite{chen2015, lee2021}.

This paper presents CleanStack, a protection mechanism that is low-overhead, highly compatible, and offers comprehensive defense against both control-flow hijacking attacks and non-control data attacks. The key idea of CleanStack is to isolate stack objects that are influenced by external user input (potentially from an attacker). By separating these tainted stack objects, CleanStack ensures that they cannot modify the return address. Additionally, CleanStack randomizes the Unclean Stack, which stores these tainted objects, to prevent modifications between them. We define this approach as Tainted Stack Object Separation (TSOS). A crucial aspect of TSOS is the identification of tainted stack objects. In this paper, we analyze both static program analysis and heuristic methods for determining these objects. Considering compatibility, we ultimately adopt the heuristic method for implementation.

We implemented CleanStack based on the LLVM compiler framework and evaluated its performance using the SPEC CPU2017 benchmark suite and real-world web server applications. To verify the effectiveness of CleanStack, we conducted statistical analysis and experimental security analysis. Specifically, in the experimental security analysis, we selected two real-world stack-based vulnerabilities, one targeting control-flow hijacking attacks and the other targeting non-control data attacks. The results demonstrate that CleanStack effectively prevents these attacks. Evaluation results show that CleanStack, by adopting a dual-stack system and randomizing stack objects in the Unclean Stack, provides effective stack protection against control-flow hijacking and non-control data attacks while incurring only a 1.73\% performance overhead.

In summary, the contributions of this paper are as follows:

Comprehensive Stack Protection: CleanStack provides robust protection against stack-based memory corruption attacks, effectively mitigating both control-flow hijacking attacks and non-control data attacks, including ROP (Return-Oriented Programming) attacks and DOP (Data-Oriented Programming) attacks.

Novel Dual-Stack System – Tainted Stack Object Separation (TSOS): We propose a new dual-stack system theory, TSOS, to enhance security by isolating tainted stack objects. Furthermore, to improve portability and applicability, CleanStack is implemented within the LLVM compiler framework.

Comprehensive Performance and Security Evaluation: To validate the performance and security of CleanStack, we applied it to the SPEC CPU2017 and Apache HTTP server benchmark suites, as well as widely used real-world applications. Experimental results demonstrate that, compared to existing stack protection techniques, CleanStack achieves an optimal balance among protection coverage, execution overhead, and deployability/compatibility, making it the most effective stack security solution available.

\section{THREAT MODEL}
\noindent This paper primarily focuses on stack-based memory corruption attacks, encompassing control-flow hijacking attacks and non-control-data attacks. The core of control-flow hijacking attacks lies in the attacker's ability to manipulate the return address, causing the program to jump to a code region that would normally be unreachable. In contrast, non-control-data attacks modify critical decision-making variables, thereby influencing the program's control flow and forcing it to execute a specific path as intended by the attacker.
We assume that when accessing stack data, a program may have stack memory security vulnerabilities, such as stack buffer overflows/underflows. Attackers can exploit these vulnerabilities to illegally read or tamper with other objects on the stack, and in extreme cases, even gain complete control over the stack memory of the process. However, attackers do not have direct access to modify the code segment. While they can leverage input-controlled stack memory corruption vulnerabilities to perform arbitrary stack memory read and write operations, they are unable to modify the code segment, as it is marked read-executable and non-writable. Additionally, attackers cannot interfere with the program loading process. These assumptions ensure the integrity of the instrumented program code at compile time and allow the program loader to securely configure dedicated segment registers for stack canaries and the unsafe stack, thereby enhancing system security. Our adversary model is consistent with those used in related research in this field~\cite{kuznetsov2014, george2024}.

\section{DESIGN}
\subsection{Overview}
The core design of CleanStack is based on the Tainted Stack Object Separation (TSOS) theory, aiming to prevent memory corruption attacks that exploit the stack by isolating and randomizing stack objects. In traditional systems, the stack grows linearly and is organized sequentially. While this structure facilitates program management, its predictability introduces severe security risks, allowing attackers to exploit deterministic memory layouts to launch non-control-data attacks and control-flow hijacking attacks. To mitigate memory corruption attacks, CleanStack adopts a dual-stack architecture, physically isolating variables influenced by external input data from regular stack objects. This prevents attackers from modifying return addresses through memory vulnerabilities. Additionally, CleanStack enforces static randomization of object placement in the unclean stack, breaking attackers' ability to predict the memory layout. The primary defense mechanisms employed by CleanStack include: 1) Tainted stack object identification – precisely identifying stack objects that may be influenced by external inputs and thus susceptible to exploitation. 2) static randomization of the tainted stack – ensuring that the placement of tainted stack objects is randomized, preventing attackers from using simple offset calculations to infer their memory addresses.

The overall workflow of CleanStack is illustrated in Figure 1, demonstrating its operational process from source code compilation to executable generation. First, the source file is compiled into an LLVM IR file using the Clang compiler with the CleanStack annotation enabled. Next, static analysis is performed to identify tainted stack objects, which are then relocated to a separate tainted stack for isolation. The layout of the tainted stack undergoes static randomization to disrupt predictable memory layouts, making exploitation significantly harder. Additionally, canary values and runtime canary checks are inserted into the code to further enhance security. Finally, the transformed LLVM IR is processed by LLVM CodeGen to generate the executable, ensuring that CleanStack's dual-stack architecture and randomization techniques effectively mitigate stack-based memory vulnerabilities.

\begin{figure}[h]
	\centering
	\includegraphics[width=3.5in]{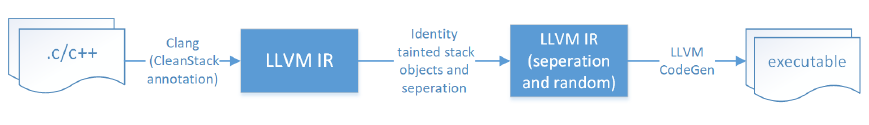}
	\caption{Illustration of CleanStack’s workflow.}
	\label{fig1}
\end{figure}

\subsection{Tainted Data Identification}
A key step in implementing CleanStack is accurately identifying stack objects that may be tainted by external inputs. External inputs (such as user-provided data, network communications, or file contents) serve as critical attack vectors. Attackers exploit program vulnerabilities by manipulating external input data to achieve malicious objectives. Therefore, identifying potentially tainted data is essential to prevent malicious exploitation.

Once external data enters a program, it can propagate throughout the system, contaminating multiple variables and creating new attack vectors. For example, an unchecked user input could be used to overwrite critical variables or modify the program’s execution path. The spread of tainted data is one of the fundamental risks in memory-related security vulnerabilities. Clean data—such as constants defined within the program scope and variables that do not depend on external input—is generally considered safe, as it remains unaffected by external influences and does not pose significant security risks. However, variables dependent on external input are vulnerable to manipulation, making the identification of dependencies between these variables a critical security measure.

To effectively identify these potentially tainted variables, CleanStack employs two methods: taint analysis and heuristic methods. Taint analysis is a widely used technique in static code analysis that tracks how data propagates within a program. It begins at external data sources (such as user inputs) and follows their propagation through the program’s control flow and data flow. This analysis helps determine the scope of tainted data propagation, the affected variables, and how they could be exploited. CleanStack uses taint labels to mark potentially dangerous data, ensuring that these tainted data objects are isolated from other safe objects in the stack. Isolating tainted data prevents it from mixing with clean data, thereby reducing the likelihood of successful exploitation. The heuristic method used for tainted data identification will be detailed in the implementation section.

The taint propagation rules in CleanStack are comprehensive, covering a wide range of operations to ensure that all potentially tainted data is identified:

1.	Assignment operation: When a tainted variable is assigned to another variable, the new variable inherits the tainted label. This is one of the basic rules of taint analysis. If external input taints a variable, any subsequent assignment involving that variable will cause the taint to propagate to the new variable. For example, if t1 is tainted and t2 = t1, then t2 will also be marked as tainted. This rule ensures that tainted data does not spread uncontrollably within the program.

2.	Unary operations: Unary operations involve manipulating a single variable. If  the operand of the operation is tainted, the result of the operation is also marked as tainted. For example, in the operation t2 = -t1, t2 inherits the taint status of t1. Unary operations commonly include negation, bitwise NOT, and absolute value calculations. CleanStack tracks these operations to ensure proper taint propagation.

3.	Binary operations: Binary operations (such as addition, subtraction, and multiplication) involve two operands. If either operand is tainted, the result of the operation is also marked as tainted. This ensures that no tainted data goes unnoticed during program execution. For example, in the expression t3 = t1 + t2, if either t1 or t2 is tainted, t3 will be marked as tainted. This rule guarantees accurate tracking of tainted data in complex arithmetic and logical operations.

4.	Comparison operations play a crucial role in program control flow. If they involve tainted data, they can influence the program’s decisions, potentially leading to vulnerabilities. By marking the result of comparison operations as tainted when one or more operands are tainted, CleanStack ensures that any decision based on tainted data is properly monitored. For example, in the comparison t3 = (u1 < t2), if t2 is tainted, then t3 will also be considered tainted. This helps prevent tainted data from causing malicious changes in control flow.

5.	Pointer arithmetic: Pointer arithmetic involves calculations on memory addresses. If a pointer or an offset in a pointer operation is tainted, the result is also marked as tainted. Pointer arithmetic is common in low-level memory management, making it a critical point for taint tracking. For example, in the operation t3 = \&v1 + t2, if t2 is tainted, t3 will inherit the taint label. This rule ensures that potential attacks exploiting pointer arithmetic (such as buffer overflows or memory corruption) are effectively mitigated.

6.	Function calls: Functions are key mechanisms for data propagation within a program, as they often accept parameters and return values. If a function receives tainted parameters, its return value is likely tainted as well. CleanStack tracks tainted data as it passes through function calls to ensure that tainted data is properly handled. For instance, in the function call t1 = f(u1), if u1 is tainted, the return value t1 will also be marked as tainted. This rule is particularly important for functions that handle input/output operations or manipulate external data, as they can easily propagate tainted data throughout the program.

These rules collectively enable CleanStack to perform comprehensive taint analysis, accurately identifying and tracking tainted data across various operations in the program. This process ensures that all potentially dangerous data is identified and isolated before it can lead to security vulnerabilities. After identifying potentially tainted stack objects, they are placed into the tainted stack to prevent these stack objects, which may be influenced by program input, from tampering with clean data such as return addresses.

\subsection{Static Randomization of the Tainted Stack}
 Once CleanStack identifies which data is tainted, the next step is to randomize the tainted stack, which stores this potentially dangerous data. Randomizing the tainted stack prevents attackers from accurately predicting the memory layout of critical data, significantly increasing the difficulty of exploitation.

Stack randomization is a widely used technique in modern security practices, and CleanStack further enhances its defense capabilities by specifically randomizing tainted data. Attackers often rely on the known memory layout of stack objects to execute attacks such as Return-Oriented Programming (ROP) or Data-Oriented Programming (DOP). To counter this threat while considering program robustness and execution overhead, CleanStack adopts static randomization, which rearranges tainted stack objects in a non-deterministic manner at the start of each program execution. This significantly increases the complexity of attacks, making them much harder to execute successfully.

In addition to stack randomization, CleanStack introduces additional defense mechanisms:

Guard Pages: CleanStack allocates guard pages alongside the tainted stack, marking them as non-accessible memory regions. If a program attempts to read from or write to these guard pages, a system error is triggered, preventing attackers from exploiting inter-stack overflows or other memory corruption vulnerabilities. Guard pages serve as a security barrier, blocking illegal memory accesses and terminating the program before an attack can escalate.

Canary Values: CleanStack also employs the stack canary technique by inserting randomly generated canary values at critical locations in the stack, such as stack frame boundaries. When a stack overflow occurs, the canary value is the first to be overwritten, allowing the program to check its integrity before returning. If the canary value has been modified, the program immediately terminates execution to prevent further exploitation. This mechanism provides an additional layer of protection against stack-based attacks.

Through the combination of these techniques—static randomization, guard pages, and stack canaries—CleanStack provides multi-layered security protection for the tainted stack, making it extremely difficult for attackers to exploit stack-based vulnerabilities. Even if one defense mechanism is bypassed, other mechanisms remain in place to effectively block the attack, ensuring system security.

As shown in Figure 2, the diagram illustrates the dual-stack architecture in CleanStack, consisting of a clean stack and a tainted stack, both growing from high memory addresses (bottom) to low memory addresses (top).

\begin{figure}[h]
	\centering
	\includegraphics[width=3.5in]{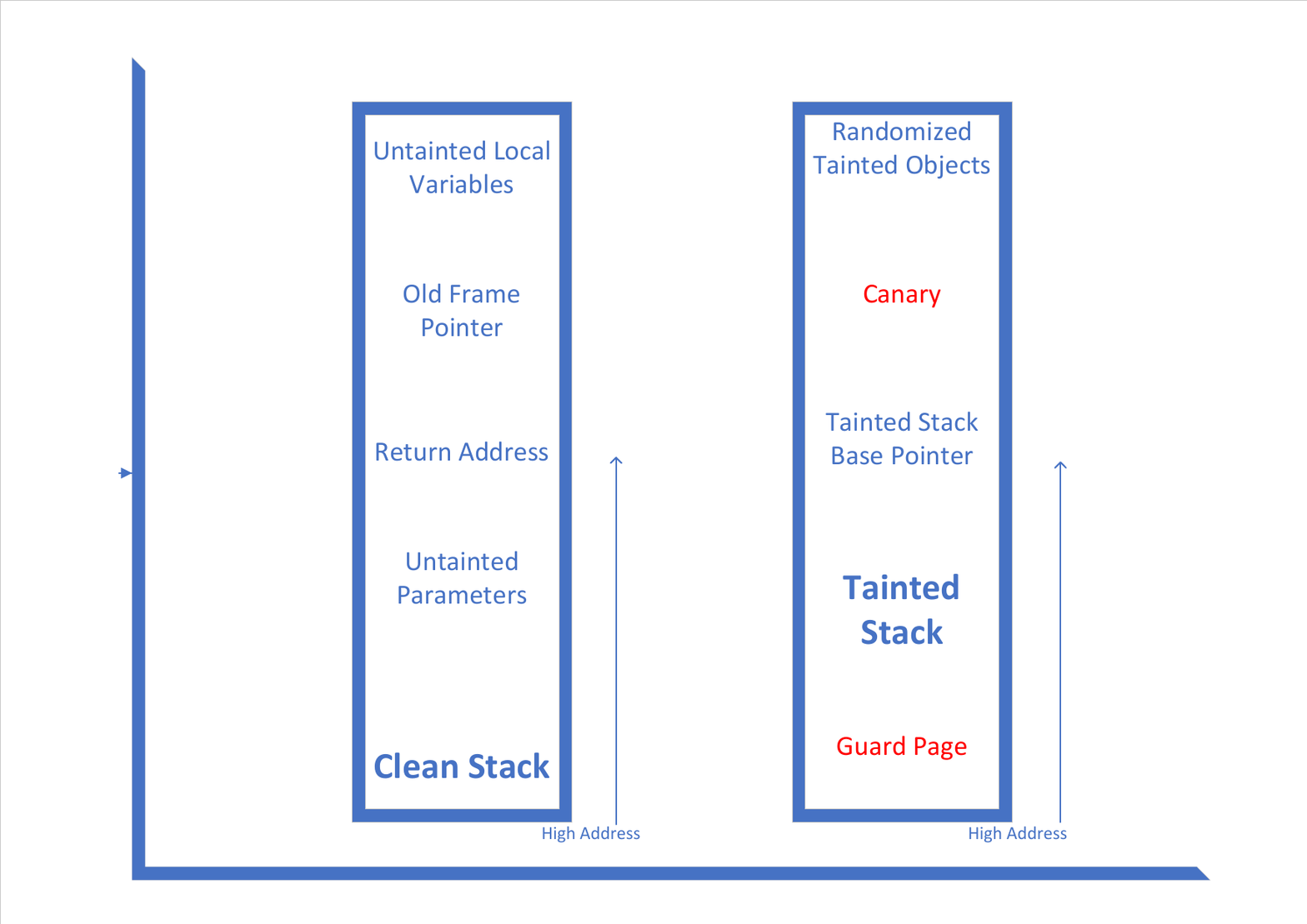}
	\caption{CleanStack’s Model.}
	\label{fig2}
\end{figure}

\subsubsection{Clean Stack (Left Side)}
\begin{itemize}
	\item \textbf{Clean Parameters}: Parameters passed to the function that have been verified as clean.
	\item \textbf{Return Address}: The address to which the function should return after execution.
	\item \textbf{Old Frame Pointer}: The saved frame pointer from the previous function call, used to restore the stack context after the function returns.
	\item \textbf{Clean Local Variables}: Local variables that have been verified to be clean by external input and are safe to use.
\end{itemize}

\subsubsection{Tainted Stack (Right Side)}
\begin{itemize}
	\item \textbf{Guard Page}: A memory protection area used to prevent stack overflow attacks between stacks.
	\item \textbf{Tainted Stack Base Pointer}: Marks the base of the tainted stack, ensuring proper memory management.
	\item \textbf{Canary}: A special value inserted into the stack to detect buffer overflow attacks, used to protect the tainted stack base pointer. 
	If this value is altered during execution, the program will terminate to prevent exploitation.
	\item \textbf{Randomized Tainted Objects}: Objects identified as potentially tainted (e.g., decision variables, data pointers). 
	These objects are arranged randomly to prevent attackers from predicting the memory layout and launching attacks.
\end{itemize}

The diagram includes short arrows on the right side of each stack, indicating the stack growth direction, which is from high address (bottom) to low address (top).
 
\section{IMPLEMENTATION}
\noindent We implemented CleanStack on the x86\_64 architecture using the LLVM/Clang-10.0.0 compilation framework on Linux-22.04. The LLVM compiler framework is used for static instrumentation of functions with the CleanStack attribute and for providing runtime library support. This section details the implementation process of the CleanStack defense system, including the identification methods for tainted stack objects (static program analysis and heuristic methods) and memory layout management using the dual-stack model.

\subsection{Identification of Tainted Stack Objects}
The identification of tainted stack objects is a core step in CleanStack, aimed at isolating stack variables potentially influenced by external inputs into a contaminated stack. Two methods are used for identifying tainted stack objects: static program analysis and heuristic methods.

\subsubsection{Static Program Analysis}
Static program analysis is a compile-time code analysis technique that tracks how external inputs propagate to stack variables~\cite{jarzabek2007, nielson2015principles}. It relies on taint analysis to identify potential tainted variables by tracing data flows. The implementation follows these steps:

To differentiate key variables storing tainted values as per the taint propagation rules outlined in section 3, a static flow-sensitive method based on an iterative data-flow analysis framework was adopted. This framework requires defining data flow equations for control flow graph (CFG) nodes, representing their entry (in) and exit (out) states. By abstractly interpreting program statements and iteratively solving equations until convergence, a stable system state is achieved.

\noindent Key components of abstract interpretation include:
\begin{itemize}
	\item \textbf{Transfer function}: Simulates instruction execution at each CFG node.
	\item \textbf{Analysis direction}: Specifies whether the analysis is forward or backward.
	\item \textbf{Join operator}: Defines how to merge attributes flowing in from predecessors or successors, such as union or intersection.
\end{itemize}

Given the granularity required by CleanStack for memory operations, each CFG node corresponds to a single instruction. The transfer function is defined as:

\begin{equation}
	f(n) = \text{gen}[n] \cup (\text{in}[n] - \text{kill}[n])
\end{equation}

where:
\begin{itemize}
	\item \(\text{gen}[n]\): Set of tainted values introduced at node \(n\).
	\item \(\text{kill}[n]\): Set of values losing taint at node \(n\).
	\item \(\text{in}[n]\): Tainted values entering node \(n\).
\end{itemize}

Taint propagation flows forward from earlier program statements, determining the direction of analysis. The union operator is chosen as the join operator, as the definition of taint is inclusive—if a variable is influenced by a legitimate path, it is considered tainted. As a forward may analysis, the data flow equations are:

\begin{align}
	\text{in}[n] &= \bigcup_{p \in \text{pred}[n]} \text{out}[p]  \quad  &  \\
	\text{out}[n] &= f(n)  \quad  & 
\end{align}

where \(\text{out}[n]\) is the set of tainted values exiting node \(n\).

To solve these equations, a worklist algorithm is employed at the function level. For inter-procedural analysis, the algorithm traverses the call graph in reverse topological order (bottom-up). This ensures that caller functions leverage the results of callee functions, enabling scalable inter-procedural taint propagation.

For functions with parameters, the algorithm is executed multiple times to track how each parameter independently propagates taint to the function’s output. During the traversal call graph, taint propagation rules are generated based on patterns in function outputs (e.g., return values and by-reference parameters). These rules describe which parameters propagate taint or whether the function itself can act as a taint source. These rules enable cross-procedural analysis of call sites during abstract interpretation.

After completing the traversal, the analysis provides stable in and out states for each instruction, revealing whether variables at specific nodes are tainted. This information identifies which stack objects in which functions may be tainted. Although the analysis is inter-procedural, it is context-insensitive, as instrumented code should remain consistent across different call contexts.

\subsubsection{Heuristic Method}
Due to compatibility issues requiring pointer analysis for the entire program, this project ultimately adopts a heuristic approach. The heuristic method identifies potential tainted variables based on predefined rules and runtime characteristics.

What types of stack objects in a function are more likely to store and propagate taint values?
\begin{enumerate}
	\item Arrays,
	\item Structures containing arrays,
	\item Other stack variables whose addresses are taken, including:
	\begin{enumerate}
		\item \textbf{Stack variables with stored addresses}\\
		When the address of a stack variable is stored in other variables or global locations, user input can directly write to these stack variables through the stored address. Address storage breaks the locality of stack variables, enabling indirect access by external code.
		
		\item \textbf{Stack variables involved in atomic operations}\\
		In multithreading environments, user input can propagate taint values across threads via shared data. Taint may overwrite variables when passed as a new value (NewValOperand) in atomic operations.
		
		\item \textbf{Stack variables whose addresses are converted to integers}\\
		When stack addresses are converted to integers (PtrToInt), their semantics become untraceable. These integers can be transmitted via files or networks, leading to indirect address leaks and propagating user input.
		
		\item \textbf{Stack variables whose addresses are passed to functions}\\
		When a stack variable's address is passed to a function, user input can write to the stack variable via that function. Insufficient boundary checks on function parameters may lead to tainted variables.
		
		\item \textbf{Stack variables whose addresses propagate via control flow}\\
		When stack variable addresses are used as return values or propagate via exception handling paths, taint values may span multiple stack frames. Addresses returned as values can be written to by other frames.
		
		\item \textbf{Stack variables accessed through complex pointer operations}\\
		Indirect access through pointer operations (e.g., typecasting, offset calculations) increases tracking difficulty. Such operations may cross variable boundaries, broadening the taint's scope.
		
		\item \textbf{Stack variables propagating via PHI nodes}\\
		When stack variable addresses propagate through PHI nodes (control flow merge points), operations on multiple paths may result in taint values overwriting original data.
	\end{enumerate}
\end{enumerate}

\subsection{Dual-Stack Memory Layout Management}
\subsubsection{Tainted Stack Object Isolation}
Tainted stack object isolation is achieved by moving susceptible stack objects to a separate unclean stack. In CleanStack, memory allocation for each susceptible stack object is adjusted as follows:

Variables are allocated on the unclean stack. This is implemented through the moveStaticAllocasToUncleanStack function, which moves statically allocated susceptible stack objects, including function arguments allocated on the stack, and the moveDynamicAllocasToUncleanStack function, which relocates dynamically allocated susceptible stack objects to the unclean stack. Additional mechanisms, such as guard pages and randomization, are employed to protect the unclean stack.

\subsubsection{Static Randomization of the Tainted Stack}
Static randomization rearranges stack objects in a non-deterministic manner during each program execution. This prevents attackers from accurately predicting the stack layout, thereby hindering potential attacks. Since the majority of memory allocations in function stacks are static, this study focuses on randomizing statically allocated memory to improve the runtime efficiency of the defense system and enhance its practicality for stack recovery points.

The computeLayout function is responsible for calculating the layout of stack objects. This function uses system hardware as a source of randomness and employs the Mersenne Twister algorithm to generate random numbers for stack object order randomization. The order of stack objects is randomized using the shuffle library function, which implements a shuffling algorithm based on the generated random numbers. This process ensures that the memory layout of static objects susceptible to contamination is randomized with each program execution.

\subsubsection{Additional Protection Mechanisms}
\begin{description}
	\item[\textbf{Guard Page}] \hspace{0pt} \\ 
	An inaccessible memory region is allocated adjacent to the unclean stack, effectively isolating the stack from other memory regions to prevent stack overflow attacks. Guard pages are allocated using \texttt{mmap} and configured as inaccessible (\texttt{PROT\_NONE}) using \texttt{mprotect}. These guard pages are typically positioned at the boundaries of the stack. During the initialization of both the main thread and new threads, guard pages are allocated to protect against stack overflows and unauthorized access. If stack operations exceed their boundaries and touch the guard page, the operating system triggers a segmentation fault (\texttt{SIGSEGV}), allowing the error to be detected and the program to be terminated.
	
	\item[\textbf{Stack Canary}] \hspace{0pt} \\ 
	A random value, known as a stack canary, is inserted at the location of the unclean stack pointer to protect it and prevent stack overflow attacks. The stack canary is implemented by loading the value of the global variable \texttt{\_stack\_chk\_guard} at the function entry and storing it on the unclean stack. Before the function returns, the canary value stored in the stack is checked against the original value. If a mismatch is detected, the function \texttt{\_\_stack\_chk\_fail} is invoked to terminate the program, preventing potential stack overflow attacks. The implementation includes loading the canary value, inserting a check logic at return points, and handling detected anomalies through conditional branches. This ensures the integrity of the unclean stack pointer and mitigates stack overflow attacks effectively.
\end{description}

\subsection{Multi-Threading Support and Non-Local Jumps}
CleanStack uses thread-local storage (TLS) to maintain independent stack pointers 
(\texttt{\_\_cleanstack\_unclean\_stack\_ptr}) and associated metadata 
(\texttt{unclean\_stack\_start}, \texttt{unclean\_stack\_size}, \texttt{unclean\_stack\_guard}) 
for each thread. This ensures stack isolation and provides each thread with its own unclean stack. 
It intercepts the \texttt{pthread\_create} function to allocate and initialize an independent 
unclean stack with a guard page for each new thread before it starts. The thread entry function 
is wrapped to initialize the unclean stack for the thread and register a cleanup function before 
invoking the thread's original start routine. 
Upon thread termination, the \texttt{thread\_cleanup\_handler} is called to release the allocated 
stack resources, preventing memory leaks.

CleanStack supports non-local adjustments (e.g., exception handling or setjmp/longjmp) by precisely managing stack pointers. During code generation, CleanStack identifies all potential stack recovery points (such as exception handling entry points and setjmp calls) and marks them as "stack recovery points." For dynamically allocated stack objects, it creates a local variable DynamicTop to store the current stack top state, updating its value whenever dynamic memory is allocated or deallocated. At stack recovery points or exception entry points, it inserts code to reload the stack pointer from DynamicTop or the static stack top, ensuring stack state consistency.

\section{Evaluation}
\noindent We evaluated the performance impact of CleanStack on the SPEC CPU2017 and Apache Bench 2.3 benchmark suites, as well as the real-world application Nginx 1.24.0. The performance evaluation experiments were conducted on a mobile workstation equipped with an Intel i7-12800HX CPU (base frequency 2.0 GHz, turbo boost 4.8 GHz, 16 GB DDR5-4800 RAM), running Ubuntu 22.04.5 with Linux kernel version 6.8. The effectiveness evaluation experiments of CleanStack were conducted on a laptop equipped with an Intel i7-4710MQ CPU (base frequency 2.50 GHz, 16 GB RAM), running in a VirtualBox 7.0.10 virtual machine environment. The host operating system was Windows 10.

\subsection{Runtime Overhead}
We compared the performance of CleanStack's dual-stack scheme with the baseline scheme that uses a conventional stack to evaluate the overhead introduced by the additional security measures. We primarily used the C and C++ benchmark programs from the SPEC CPU2017 benchmark suite (running three iterations on the x86-64 platform). The runtime overhead evaluation was conducted on an Intel Core i7-12800HX processor running Ubuntu. On Ubuntu, we used the compiler-rt runtime library with dual-stack support. In all evaluations, we used binaries compiled natively with LLVM 10.0.0 as the baseline and enabled the compiler flag -fsanitize=clean-stack for CleanStack's dual-stack scheme.

\begin{table}[h]
	\centering
	\caption{Performance Overhead of CleanStack on the x86-64 ISA}
	\label{tab:cleanstack_performance}
	\begin{tabular}{|l|c|c|}
		\hline
		\textbf{Benchmark} & \textbf{Baseline} & \textbf{CleanStack Overhead} \\
		\hline
		500.perbench\_r & 499  & 3.41\%  \\
		502.gcc\_r      & 633  & 0.16\%  \\
		505.mcf\_r      & 1178 & -0.59\% \\
		520.omnetpp\_r  & 1351 & 2.44\%  \\
		523.xalancbmk\_r & 583  & 1.72\%  \\
		525.x264\_r     & 311  & 3.54\%  \\
		531.deepsjeng\_r & 383  & 0.52\%  \\
		541.leela\_r    & 599  & 5.34\%  \\
		557.xz\_r       & 756  & 0.93\%  \\
		\hline
		\textbf{Mean}   & -    & \textbf{1.73\%} \\
		\hline
	\end{tabular}
\end{table}

We summarized the SPEC test results in Table~\ref{tab:cleanstack_performance} (Performance Overhead). As shown in the table, the average performance overhead of CleanStack is 1.73\%, significantly lower than the 22\% runtime overhead introduced by StackArmor and the 14\% overhead introduced by SaVioR. 

Among the nine benchmark tests, \textbf{541.leela\_r} exhibits the highest runtime overhead due to several reasons. First, it employs Monte Carlo Tree Search (MCTS), which involves deep recursive calls where each recursive function may contain multiple local variables. If these variables are influenced by user input, CleanStack moves them to the unclean stack, introducing additional stack management overhead. Each recursion requires accessing \texttt{UncleanStackPtr} for stack operations, resulting in extra store and load instructions. Additionally, if the function contains \texttt{stackrestore}, the extra \texttt{stackrestore} processing in \texttt{setjmp/longjmp} scenarios increases unnecessary storage and restoration operations. 

Second, \textbf{541.leela\_r} involves a large amount of floating-point computation, and the unclean stack disrupts the alignment of floating-point data, negatively impacting floating-point computation performance. 

Third, its data access pattern is highly randomized, leading to an increased number of pointer dereferences and a higher cache miss rate. 

In contrast, another AI search benchmark program, \textbf{531.deepsjeng\_r}, exhibits significantly lower runtime overhead (0.52\%) as it employs Alpha-Beta pruning, which reduces recursive calls, focuses computations on integer operations with less dependency on memory access, and follows a more predictable data access pattern, resulting in a higher cache hit rate.

Notably, the \textbf{505.mcf\_r} benchmark performs better in the CleanStack version than in the original implementation. As a memory-intensive benchmark, most of its execution time is spent in loops that continuously access large arrays on the heap. CleanStack improves performance by transferring function local variables that may be influenced by external input to the unclean stack, while keeping other safe variables in the clean stack, placing them closer to the return address and callee-saved registers. This effectively enhances the locality of frequently accessed stack data, thereby optimizing performance.

To evaluate CleanStack's performance in I/O-intensive applications, we measured the throughput of a CleanStack-protected real-world application, the Nginx Web Server, using the Apache HTTP server benchmarking tool. The benchmark simulated eight worker processes handling 500 to 1000 Keep-Alive requests, with each request involving 5 to 50KB files, and was repeated across 11 test rounds to ensure result stability and reliability. As shown in Table II, the results indicate that CleanStack improves the performance of the Nginx Web Server. This improvement is attributed to two key factors. First, Nginx functions are relatively large with fewer transitions between function calls [16], making the overhead of setting up additional stack frames negligible in comparison to the overall function execution time. Second, the objects moved to the unclean stack are typically large arrays and shifting them out of the clean stack enhances the locality of frequently accessed stack values, such as temporarily stored CPU register values, return addresses, and small local variables unaffected by external input. This, in turn, reduces cache misses and improves overall performance.

\begin{table}
	\begin{center}
		\caption{Performance Impact of CleanStack on Nginx Web Server Throughput}
		\label{tab:nginx_performance}
		\begin{tabular}{|c|c|c|}
			\hline
			Request-File Size (Clients-KB) & Baseline & CleanStack Overhead \\
			\hline
			1000-5  & 110889.33 & -2.31\% \\
			500-5   & 109661.15 & -3.30\% \\
			1000-10 & 110889.33 & -4.96\% \\
			500-10  & 109661.15 & -3.58\% \\
			1000-20 & 110889.33 & -2.31\% \\
			500-20  & 109745.39 & -2.11\% \\
			1000-50 & 109962.61 & -1.61\% \\
			500-50  & 109661.15 &  0.00\% \\
			\hline
			Mean    &           & -2.52\% \\
			\hline
		\end{tabular}
	\end{center}
\end{table}

\subsection{Memory Overhead}
We evaluated Smokestack’s memory overhead by measuring the maximum resident set size (ru\_maxrss) during the execution of the SPEC 2006 benchmark suite. As shown in Table III, the results indicate that the average memory overhead is only 0.04\%. Interestingly, 541.leela\_r, which has the lowest memory overhead (even lower than the baseline), exhibits the highest performance overhead. This is because 541.leela\_r relies on a neural network for position evaluation, involving extensive floating-point computations. The baseline scheme may have used a conservative memory alignment strategy, introducing additional padding, whereas objects stored in the unclean stack are not explicitly aligned, thereby reducing wasted memory space. As a result, 541.leela\_r shows the lowest memory overhead while incurring the highest performance overhead.

\begin{table}
	\begin{center}
		\caption{Memory Overhead Evaluation (ru\_maxrss).}
		\label{tab:memory_overhead}
		\begin{tabular}{| l | r | c |}
			\hline
			Benchmark & Baseline (KB) & CleanStack Overhead \\
			\hline
			500.perbench\_r    & 205644  & 0.97\%  \\
			502.gcc\_r         & 1136992 & 0.17\%  \\
			505.mcf\_r         & 624020  & 0.00\%  \\
			520.omnetpp\_r     & 247728  & 0.25\%  \\
			523.xalancbmk\_r   & 491084  & 0.23\%  \\
			525.x264\_r        & 159936  & 0.00\%  \\
			531.deepsjeng\_r   & 718080  & 0.00\%  \\
			541.leela\_r       & 121404  & -1.27\% \\
			557.xz\_r          & 794372  & 0.03\%  \\
			\hline
			Mean               & --      & 0.04\%  \\
			\hline
		\end{tabular}
	\end{center}
\end{table}

\subsection{Security Analysis}
Stack-based Use-After-Free (UAF) vulnerabilities are rarely reported [28]. Most UAF attacks target heap-allocated objects because attackers can explicitly manage heap memory. Therefore, this section primarily discusses the effectiveness of CleanStack against spatial attacks. To evaluate CleanStack's defense capabilities, we conducted both statistical analysis and empirical analysis. The statistical analysis assessed the theoretical security of CleanStack by calculating the probability of an attacker successfully executing a memory corruption attack. The empirical analysis, on the other hand, tested CleanStack’s protection effectiveness in real-world scenarios. For this purpose, we integrated CleanStack into two real-world applications, Nginx and DNSTracer, and conducted control-flow hijacking attacks and non-control data attacks against two known CVE vulnerabilities to evaluate its effectiveness.

\subsubsection{Statistical Security Analysis}
Since CleanStack’s unclean stack applies randomization to tainted stack objects as a probabilistic protection mechanism, it helps mitigate spatial memory corruption attacks. Therefore, we present the probability of an attacker successfully executing a spatial memory corruption attack.

Spatial attacks are a type of memory corruption attack that exploit out-of-bounds array access, pointer arithmetic errors, or buffer overflows to read or write data in unauthorized memory regions. The key characteristic of such attacks is that they access memory addresses not correctly allocated to variables, thereby compromising data integrity and, in some cases, even taking control of the program’s execution flow.

Spatial attacks can be categorized into contiguous spatial attacks and non-contiguous spatial attacks. Contiguous spatial attacks overwrite the entire memory region between a vulnerable variable (e.g., a buffer) and a target variable (e.g., a return address or a decision-making variable), causing all intermediate variables to be corrupted. A classic example is the buffer overflow attack, where an attacker writes beyond a buffer's allocated memory, overwriting adjacent stack variables, including control-flow data. In contrast, non-contiguous spatial attacks precisely modify the target variable without affecting intermediate variables. These attacks are harder to detect because they do not disrupt an entire memory region and can bypass defense mechanisms such as Stack Canary. A typical example is the format string attack, where an attacker exploits vulnerabilities in functions like printf() to manipulate arbitrary memory locations with pinpoint accuracy.

These two types of spatial attacks on the stack can be further classified into intra-frame and inter-frame attacks. Contiguous inter-frame attacks are typically mitigated by Stack Canary and Guard Page protections. However, the probability of successfully executing a contiguous intra-frame attack, non-contiguous intra-frame attack, or non-contiguous inter-frame attack depends on two key factors: (1) the proportion of functions affected by external user input (i.e., malicious input) relative to the total number of functions in the program and (2) the number of randomized unclean stack objects. The probability of a successful attack can be expressed as:(unclean stacks/total functions)*(unclean stacks/unclean allocas). In the SPEC CPU2017 benchmark, the probability of a function being influenced by external input is 18.99\%, and the average number of unclean stack objects per function is 2.56, resulting in an overall attack success probability of 7.44\%. If the insertion position of the Stack Canary is further randomized, the success probability drops even lower, reducing by a factor of 0.5 to 3.72\%.

Limitations. It is important to note that the randomization approach used in CleanStack is static randomization, ensuring that runtime overhead remains within a deployable range. However, this introduces potential limitations. If an attacker gains knowledge of the unclean stack layout after compilation or, alternatively, if the attacker does not initially know the unclean stack layout but can attempt multiple attacks, CleanStack’s defense effectiveness may be weakened.

\subsubsection{Empirical Security Analysis}
We conducted an effectiveness evaluation on two CVE vulnerabilities to assess their exploitability under the CleanStack protection mechanism. First, we performed experiments in a VirtualBox virtual machine running Ubuntu 13.04, where we manually analyzed and successfully exploited the vulnerability to execute a Control-Flow Hijacking attack. Second, we reproduced a Data-Oriented Programming (DOP) attack using the DOP attack tool proposed by Hu et al., conducting the experiment in a VirtualBox virtual machine running Ubuntu 12.04. Based on these experiments, we further evaluated the exploitability of these vulnerabilities when CleanStack protection was enabled.

In dnstracer 1.9 and earlier versions, a stack-based buffer overflow vulnerability exists, allowing an attacker to construct an overly long command-line argument that triggers an improper strcpy call on argv[0], leading to denial of service (DoS) (application crash) or even the execution of malicious code. The root cause of this vulnerability is the lack of length validation when copying argv[0] into the array argv0. When argv[0] exceeds 1024 bytes, a buffer overflow occurs, corrupting stack data, including the return address. In our experiment, we simulated an attacker's role by constructing an overly long command-line argument, causing an overflow in argv0 and successfully overwriting the return address, hijacking the program’s control flow. This allowed execution to jump to attacker-controlled shellcode, ultimately executing system("/bin/sh") and opening an interactive shell. However, in the CleanStack-hardened version of dnstracer, CleanStack successfully prevented this attack. The defense mechanism relies primarily on Tainted Stack Object Separation, where CleanStack allocates vulnerable local variables like argv0 to a separate unclean stack, ensuring that it does not share the same stack region as the return address. As a result, even if argv0 overflows, it cannot overwrite the return address, effectively preventing control-flow hijacking by the attacker.

In ProFTPD 1.3.0 and earlier versions, the sreplace function contains a stack-based buffer overflow vulnerability, which was successfully exploited by Hu et al. to demonstrate a Data-Oriented Programming (DOP) attack, ultimately achieving the extraction of OpenSSL private keys.

ProFTPD stores the OpenSSL private key in a buffer and accesses it through an eight-level pointer chain, where the only non-randomized pointer is the base address pointer. To successfully extract the private key, an attacker must de-randomize the seven global pointers using memory leakage techniques. By leveraging multiple rounds of DOP gadget chains, Hu et al. managed to bypass ASLR and extract the OpenSSL private key.

The core of this attack exploits a memory corruption vulnerability in sreplace(), where the attacker continuously overwrites the loop counter, using it as a DOP dispatcher gadget to link virtual instructions and construct a complete DOP attack chain.

Under CleanStack protection, this DOP attack was successfully prevented. CleanStack mitigates the attack by moving buffers (including the overflow buffer) to the unclean stack and applying randomization. This ensures that the overflow buffer and the variable cp in the assignment gadget are placed in separate stack spaces, preventing the attacker from modifying cp through memory corruption. As a result, the assignment gadget fails, which is a crucial component in linking the dispatcher gadget and other key gadgets. Since the assignment gadget is rendered ineffective, the entire DOP attack chain collapses, ultimately leading to a failed attack.

\section{Related Work}
\noindent The concept of StackGuard~\cite{cowan1998}, introduced by Cowan et al. in 1998, laid the foundation for stack protection by specifically targeting buffer overflow attacks. StackGuard inserts a "canary" value before the return address on the stack and checks its integrity before the function returns; if this value is altered, a buffer overflow attack is suspected, and the program enters a fail-safe state. This method effectively thwarts stack smashing attacks, where an attacker overflows a buffer to overwrite the return address and redirect execution to malicious code. However, StackGuard has limitations, particularly when facing sophisticated attacks that can bypass or even predict the canary value.

Shadow stacks~\cite{ave2000, chiueh2000} represent an evolution in stack protection that enhances control-flow integrity. Traditional shadow stacks maintain a separate stack for return addresses, which is checked against the main stack to ensure they match upon function return. Dang et al.~\cite{dang2020}found that traditional shadow stacks typically incur a performance overhead of around 10\% due to additional memory operations. To reduce this overhead, the authors proposed parallel shadow stacks, which use fixed offsets from the main stack, bringing down the overhead to approximately 3.5\% and simplifying the lookup process. Despite these performance improvements, shadow stacks still face challenges related to concurrency and potential exploitation by timing attacks.

SafeStack~\cite{kuznetsov2014}, introduced as part of the LLVM compiler suite, employs a dual-stack approach to stack protection by dividing stack objects into a safe stack and a regular stack, depending on whether they are likely to be targets of attacks. Determine whether an object is safe based on whether the accessed memory exceeds the allocated memory range, and place objects that cannot be proven safe on the unsafe stack. This separation aims to minimize the risk of buffer overflows impacting critical control data, such as return addresses and function pointers. While SafeStack reduces the likelihood of control-flow hijacking, it does not directly address data-oriented attacks, which primarily involve manipulating non-control data on the stack.

Smokestack~\cite{aga2019} uses a runtime stack layout randomization method to defeat DOP attacks. The principle is to full permutation the locations of all variables on the stack of a function. For example, there are four variables on the stack, then the stack layout after full permutation has a total of 4! (=24) types. At runtime, one of the stack layouts is chosen based on the generated true random number. In addition, to prevent this protection method from being bypassed by attackers, they set a label at the prologue of each function and check whether the function label matches at epilogue. This method of randomly selecting a stack layout every time a function is called makes it impossible for a DOP attacker to know the relative distance of objects on the stack and thus cannot construct an effective attack. However, this method of randomizing the stack layout at runtime necessarily results in excessive performance overhead, especially for large programs.

StackArmor~\cite{chen2015} is a comprehensive stack security protection technique designed for binary programs, specifically targeting stack-based memory error vulnerabilities. By utilizing binary analysis and rewriting techniques, StackArmor significantly reduces the spatial and temporal predictability of traditional call stack structures, effectively preventing attackers from exploiting stack-based vulnerabilities. Compared to traditional protection schemes, StackArmor does not require access to source code to defend against various stack attacks and employs a policy-driven security mechanism that allows users to flexibly balance security and performance. However, its drawbacks include a high runtime overhead of up to 22\% and a lack of support for C++ exception handling, making it impractical for real-world deployment.

SaVioR~\cite{lee2021} employs stack layout randomization to prevent attackers from predicting the locations of vulnerable objects. This approach introduces random padding and distributes stack frames across multiple regions, effectively mitigating both spatial and temporal attacks. While SaVioR demonstrates strong security performance, it incurs a runtime overhead of approximately 14\%, which may impact high-performance applications. Additionally, it can alter the representation of pointers in memory, posing challenges for practical deployment.

OPTISAN~\cite{george2024} is a stack spatial memory defense system designed to maximize the number of protected stack objects within a given cost budget. It analyzes programs to identify potentially exploitable stack objects and estimates the overhead of various defense operations, such as metadata management and spatial safety checks, to provide flexible defense strategy selection. OPTISAN employs a Mixed-Integer Non-Linear Programming (MINLP) formulation to generate an optimal defense layout and integrates Baggy Bounds~\cite{akritidis2009} and AddressSanitizer~\cite{serebryany2012} to increase the number of protected memory operations by 18.4\%. Under cost budget constraints, OPTISAN can flexibly apply multiple defense mechanisms to effectively prevent the exploitation of stack spatial memory errors. However, since OPTISAN does not consider the protection of global variables, stack objects that are both global and stack-based remain vulnerable to attacks.

\section{Concusions}
\noindent We propose CleanStack, a stack security protection mechanism that integrates isolation mechanisms and randomization strategies, designed to safeguard C/C++ applications from stack-based memory corruption attacks. To achieve this, CleanStack employs an heuristic approach to identify stack objects in the target program that are influenced by external user input. These tainted stack objects are then isolated into an unclean stack, and their positions within the unclean stack are randomized. Our statistical analysis and experimental security evaluation demonstrate that CleanStack can effectively mitigate control-flow hijacking attacks and non-control data attacks with high probability. Additionally, we have integrated CleanStack into the SPEC CPU2017 and Apache HTTP server benchmark suites and evaluated its performance and compatibility in real-world applications, including the Nginx server. The experimental results further confirm that CleanStack is practically feasible and deployable, making it suitable for various architectures and operating systems. CleanStack is available at https://github.com/leichong98/CleanStack/

\balance

\begin{IEEEbiography}[{\includegraphics[width=1in,height=1.25in,clip,keepaspectratio]{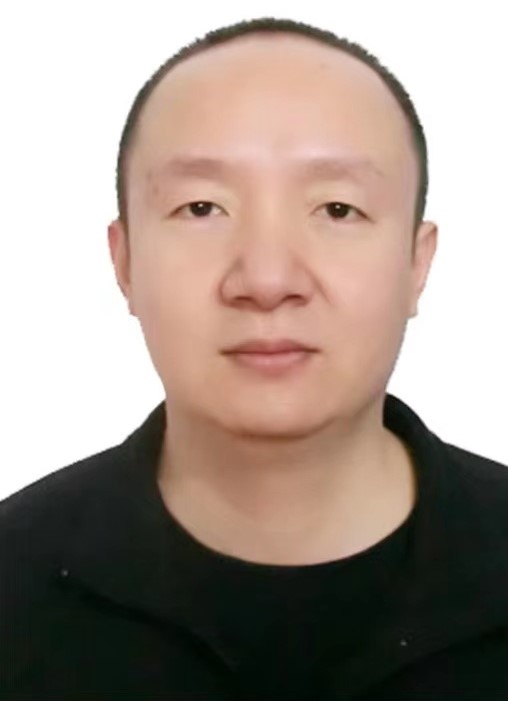}}]{Lei Chong}
	received his BS degree in automation from Wuhan Textile University, Wuhan, China, in 2005,
	and MS degree in control theory and control engineering from Guangdong University of Technology, 
	Guanzhou, China, in 2010. Currently, He is a lecturer at Sichuan University of Arts and Science. His research interests include system security and software security. E-mail: 20170005@sasu.edu.cn
\end{IEEEbiography}

\end{document}